\documentclass[12pt]{article}
\usepackage{epsfig,graphicx}

\title{Magnetic confinement and the Linde problem   }
\author{  Yu.A.Simonov,\\ State Research
Center\\Institute of Theoretical and Experimental Physics\\ 117218, Moscow,
B.Cheremushkinskaya 25, Russia}
\newcommand{\be}{\begin{equation}}
\newcommand{\ee}{\end{equation}}

\def\la{\mathrel{\mathpalette\fun <}}

\def\fun#1#2{\lower3.6pt\vbox{\baselineskip0pt\lineskip.9pt
\ialign{$\mathsurround=0pt#1\hfil ##\hfil$\crcr#2\crcr\sim\crcr}}}

\newcommand{{\SD}}{\rm SD}

\newcommand{\vex}{\mbox{\boldmath${\rm x}$}}
\newcommand{\vey}{\mbox{\boldmath${\rm y}$}}

\newcommand{\vep}{\mbox{\boldmath${\rm p}$}}
\newcommand{\veb}{\mbox{\boldmath${\rm b}$}}

\newcommand{\vez}{\mbox{\boldmath${\rm z}$}}

\newcommand{{\Mc}}{\mathcal{M}}

\newcommand{\lan}{\langle}
\newcommand{\ran}{\rangle}

\begin{document}

\maketitle

\begin{abstract}
Perturbation theory of thermodynamic potentials in QCD at $T>T_c$ is considered
with  the nonperturbative background vacuum taken into account. It is shown
that the colormagnetic confinement in the QCD vacuum prevents the infrared
catastrophe of the perturbation theory present in the case of the  free vacuum
(the ``Linde problem''). A short  discussion is  given  of the applicability of
the nonperturbative formalism at large $T$ and of the relation with the HTL
theory. The observation of A.D.Linde, that the terms $O(g^n), n>6$ contribute
to the order $O(g^6)$ is confirmed also with the account of the colormagnetic
confinement, and it is shown that the latter makes these terms IR convergent,
and summable. As a result one obtains the  selfconsistent theory of the gluon
plasma.

\end{abstract}

\newpage

\section{Introduction }

 The Linde problem in the thermal QCD has a long history, see \cite{10,11} and
 refs therein. It has occurred in the thermal perturbative QCD, where similarly
 to QED, the originally massless  constituents (gluons) acquire effective
 perturbative mass operators $m(T)$, which regulate the  convergence of $g^n$
 terms and of the whole perturbative series. Correspondingly, the
 colorelectric screening mass $m_D(T)$,obtained from $\Pi_{00} (T)$ (similarly
 to the QED case) starts from $gT$,  however the  colormagnetic screening mass
 does not exist perturbatively  \cite{10,11} (again, as in QED),  and if
 introduced effectively as $O(g^2 T)$, the perturbative series is not defined
 at the order $g^6 $ (problem (1)). Linde also remarks, that  the higher order diagrams
 contribute to the same order (problem (2)).

 Meanwhile the effective perturbative theory of thermal QCD (the
 hard-thermal-loop  (HTL) theory) was developed in \cite{13,14}, using the
 colorelectric $m_D (T)$ and the resummation technic through order $g^2,g^3,
 g^4, g^5$, which appears to be quite successful, see \cite{5a} for a review. The
 nonperturbative nature of the magnetic scale $g^2T$, which appears necessarily
 at $O(g^6)$ can be connected to the 3d Yang Mills theory, see e.g. \cite{6a}.

 A natural question arises, how this situation can be explained and treated in
 a 4d approach to QCD, where nonperturbative (np) physics (including
 confinement) is taken into account?

 In what follows we shall consider the np approach to QCD, developed in
 \cite{1,2,3,4,5,6}. For  an alternative approach see \cite{13a,14a}.

Most effects in QCD at low and intermediate energies cannot be  explained
without  np dynamics, which enters in the  theory, e.g. via  the string tension
$\sigma$, or a mass of some meson $(\rho, K$), or else the constant
$\Lambda_{QCD}$, entering in $\alpha_s (Q)$. The approach of the np vacuum,
ensuring confinement, and stabilizing perturbation theory, was developed in
\cite{1}  for QCD at zero temperature $T$, and in \cite{2,3} for $T>0$, see
reviews \cite{4,5} for $T=0$
 and \cite{6} for $T>0$. The problem of the confinement and
  deconfinement is treated in our approach, called the Field
Correlator method (FCM), taking into account two kinds of colorelectric
correlators $D^E(z), D_1^E(z) \sim \lan E_{i} (x) E_{i}(y)\ran$, and  two kinds
of colormagnetic, $D^H(z), D_1^H(z)\sim \lan H_i (x) H_i(y)\ran$, where the
first ones $(D^E, D^B)$ are of purely nonabelian character, while $D_1^E,
D_1^H$ exist also in QED.

Assumed in \cite{2} and later  confirmed on the lattice \cite{9a}, that the
nonabelian colorelectric correlator $D^E(z)$ vanishes together with confinement
at $T_{c }$, while all others stay nonzero for $T>T_c$, in particular the
nonabelian colorelectric correlator $D^E_1$, responsible for the nonzero
Polyakov lines, while  the nonabelian colormagnetic correlator $D^H(z)$ ensures
the magnetic confinement for the motion in the spatial planes. This property
was studied in the FCM formalism in \cite{7}, and analytically and numerically
in a different approach in \cite{8}, see also \cite{9} for later developments.

As  a result of magnetic confinement there appears the spatial string tension,
which defines the area law of the spatial projection of any Wilson loop in 4d,

\be \lan W(C)\ran = \exp \left(-\sigma_s A_{3d} (C)\right), ~~  \sigma_s
=\frac12 \int d^2 z D^H(z).\label{1a}\ee

 Moreover, $D^H(z)$ can be  calculated
via the gluelumps \cite{23}, known both analytically \cite{17*} and on the
lattice \cite{24}, which yields the relation

\be \sqrt{\sigma_s (T)} = c_\sigma g^2 (T) T, \label{2a}\ee where  $c_\sigma$
is of the np origin, as shown in the appendix.   This coincides with the
lattice data results \cite{16}, where   $ c_\sigma = 0.566\pm 0.013.$

The two-loop approximation is generally used for  $g^2(T)$ \cite{9*}.

One can now consider any QCD diagram and the whole perturbative series as being
immersed in the np vacuum, so that all closed loops in the 3d space are covered
by the confining film, while for  4d loops covered is its  3d projection. This
fact turns over the whole thermal QCD dynamics.

 Namely, as we show below,  the
spatial Wilson loops not only serve as a cut-off factor at the distance $X_{\rm
max} \sim \frac{1}{\sqrt{\sigma_s}}$, but due to Eq. (\ref{2a}) this cut-off
depends on $g(T)$ and  converts  the perturbative $O(g^n)$ term into $O(g^6)$.
(Problem 2 of Linde). Exactly the  same situation would occur, if instead of
spatial confinement one introduces  the magnetic mass of gluons.

Then  again Eq. (\ref{2a}) implies that the magnetic mass $m^H_D \sim
\sqrt{\sigma_s}$ as a cutoff parameter   makes the  4 loop integral convergent,
which resolves, what can be called the problem 1 of Linde, as will be clarified
below. This is purely np result, irrespectively of the appearing $g^2$ factor.
The problem (2) of Linde (see (iii) in  \cite{10} on p. 290) that  the sum of
the infinite ladder of gluon loops with $n>4$ contributes to the  same order
$g^6T^4$  also occurs in this case of magnetic mass.

However,  the notion  of magnetic mass (or any other effective mass) is
irrelevant in the case of  confinement, since gluons are connected by the
confining string, which constitutes the most part of the total energy (mass) of
the system in  contrast to  the free motion of a gluon with any effective mass.


Coming back to the results of the perturbation theory  and
 comparing HTL results with the lattice calculations , one can conclude, that
 the  $O(g^6)$ term is basically important for $T\leq 0.5$ GeV (see e.g. Fig.
 1 of \cite{6a}). As a result a
 new HTL version appeared in \cite{25**}, called ``the $O(g^6)$ fitted'' HTL
 contribution, as well as ``the $O(g^6)$ fitted $ +$ nonpert.'' version. As we shall
 show below, the $O(g^6)$ terms indeed contain the whole series $O(g^n), n>6 $,
 as was shown by Linde \cite{10}, but in addition the colormagnetic confinement
 makes these terms finite and summable. All this makes our analysis and
 discussion of the Linde problems even more timely and relevant.

 The paper is organized as follows. In the next section we describe
 qualitatively the possible solution of the problem, in section 3 we
  write the general
 background field formalism for  the thermodynamic potential, and define its
 perturbation series and
we study the gluonic multiloop diagram with spatial (magnetic)
 confinement. We define its infrared and ultraviolet properties, showing that
 indeed the presence of $\sigma_s$ prevents the IR divergence of any diagram.

 Section 4 is devoted to the summary and prospectives.

 In appendix a detailed derivation is given of $\sigma_s$ in terms of gluon
 propagators and gluelumps.

 \section{Qualitative analysis of the  perturbative diagrams}

Coming back to the Linde problem (\ref{1a}), the standard perturbation theory
(without nonperturbative background), which proceeds essentially in 3d, becomes
infrared divergent, starting with the 6-th order in $g$ \cite{10,11}.  In
essence, the problem occurs due to very weak
 fall-off of   the gluon propagator in 3d without $\sigma_s$, e.g. in the $x$-space \be G(x,y) \sim
\frac{T}{ \pi |\vex-\vey|}, ~~ |\vex-\vey| \gg 1/T\label{3a}\ee

Let us now consider a $n$-th order diagram of the thermal perturbation theory,
an example of this diagram for $n=8$ is shown in Fig. 1. One can count the
number of gluon propagators, Eq. (\ref{3a}), in  the  diagram:$ N_{\rm prop} =
\frac{3n}{2}$.

The number of vertices with  derivatives $\frac{\partial}{\partial x_i}$ at
each vertex is  $n$, the  number of space integrals $\frac{d^3 x^{(i)}}{T}$  in
each vertex is $n$, however one integral yields the overall volume, so that the
amplitude can be written as \be A_n \sim g^n \prod^n_{i=1} \int \frac{
d^3x^{(i)}}{T} \frac{\partial}{\partial x^{(i)}} \prod^{N_{\rm prop}}_{k=1}
G^{(k)}(x^{(i)}-x^{(j)}) \sim  \frac{V_3}{T} \bar A_n.\label{a4}\ee

As a result one  obtains the  spatial dimension of the amplitude $\bar{  A}_n$
in terms of an  overall  upper limit of 3d coordinate $X$,

\be \bar A_n = g^n X^{\frac{n}{2} -3} T^{\frac{n}{2}+1}.\label{a5}\ee

It is clear from (\ref{a5}), that $\bar A_n$ is  IR divergent for $n\geq 6$, in
agreement  with the Linde problem 1 \cite{10}.

Now let us take into account the spatial (colormagnetic), confinement in 3d,
which can be introduced in (\ref{a4}) in the form of the area law factor $\lan
W(C) \ran = \exp (- \sigma_s S_{\min})$, as in (\ref{1a}) with the minimal
surface $S_{\min} = A_{3d} (C)$, covering all diagram in Fig.~1. For  us it is
only important, that $S_{\min}$ is quadratic in coordinates $x^{(i)}, x^{(j)}$,
and consequently it behaves at large $X$ as $S_{\min}\sim X^2$. Being positive
definite, it makes the $\lan W(C)\ran$ the real cut-off function, as it was
proposed in \cite{12}, and  can make the spatial integrals to converge, namely
\be \bar A_n^{(\rm conf)} = g^n T^{\frac{n}{2}+1} \int (dX)^{\frac{n}{2}-3}\exp
(-\sigma_s |X|^2).\label{a6}\ee

It is clear, that introducing the dimensionless coordinate $Y=\sqrt{\sigma_s}
X$, one obtains the following representation for $\bar A_n^{(\rm conf)}$. \be
\bar A_n^{(\rm conf)} = g^n T^{\frac{n}{2}+1}
(\sqrt{\sigma_s})^{-\left(\frac{n}{2}-3\right)}J_n\label{a7}\ee where $J_n$ is
a dimensionless converging integral.

Now taking into account Eq. (\ref{2a}), $\sqrt{\sigma_s} \sim g^2(T) T$, one
obtains finally \be \bar A_n^{(\rm conf)} = g^6 T^4 C_n, ~~ C_n = J_n
(c_\sigma)^{3-\frac{n}{2}}\label{a8}\ee

Eq. (\ref{a8}) exemplifies the second part of the Linde problem: all the series
with  $n\geq 6$ contributes to the $O(g^6)$ term.

Note, that we have  not introduced above the magnetic or  any other mass
parameters for gluons, since in the case of confinement  the notion of mass can
be ascribed only to the given string state, containing two (or more) gluons,
connected by the adjoint string.

Nevertheless, if we introduce the effective   mass of the gluon, $m_{\rm
mag}(T)$, then  the gluon Green's function acquires an additional factor $\exp
(- m_{\rm mag}|\vex - \vey|)$, and these factors can be  assembled in the total
factor
 $\exp (-m_{\rm mag} \sum_{i,j} |\vex_i-\vex_j|)$, which would replace $\exp (-
 \sigma_s |X|^2)$ in (\ref{a6}).

 As a result one obtains instead of (\ref{a7}) the  representation
 \be \bar A_n^{(\rm mass)} = g^2 T^{\frac{n}{2} +1} (m_{\rm
 mag})^{-\left(\frac{n}{2} -3\right)}J_n^{(\rm mass)}, \label{a9}\ee
 and assuming for    $m_{\rm mag}$ the form of magnetic mass  $m_{\rm mag} =
 c_m g^2 T$, one again comes to  the  result (\ref{a8}). In this way one
 obtains that both spatial confinement and magnetic mass yield the same
 qualitative result: the sum of all $g^n$ terms with $n\geq 6$   contributes to
 the $O(g^6)$ term, in agreement with the second problem of Linde \cite{10},
  and in both cases the space integrals converge.
 In the next section we shall make our arguments more concise, developing
 a special representation   for   a  4-point (or 3-point) diagram
 with confinement taken into  account.

\section{Background perturbation theory in magnetic confinement}

In  this section we exploit the background perturbation theory, developed in
\cite{2,3}, to study soft and hard regimes of the internal integrations and to
demonstrate the role, which is played in this process by the  magnetic
confinement. Since we are mostly interested in the high $T$ gluon
contributions, we  confine ourselves to the case of pure gluodynamics.

We split the gluonic field $A_\mu$ into nonperturbative (np) background $B_\mu$
and the perturbative part $a_\mu$

\be A_{\mu}=B_{\mu}+a_{\mu} \label{a1}\ee and  the partition function $Z$ can
be written as a double average, using 'tHooft identity \cite{2,3} \be
 Z  \equiv<<\exp(-S(B+a))>_a>_B \label{2}\ee
 where the action $S$  contains the standard gluon, ghost and gauge fixing terms
 and in particular the triple vertices $a^3, a^2B$ .

 The inverse gluon propagator can be written as

 \begin{equation}
G^{-1} =- D^2(B)_{ab} \cdot \delta_{\mu\nu} - 2 g F^c_{\mu\nu}(B)
f^{acb}\label{3}
\end{equation}
where
\begin{equation}
(D_{\lambda})_{ca} = \partial_{\lambda} \delta_{ca} - ig T^b_{ca}
B^b_{\lambda}.\label{4}
\end{equation}

In what follows we shall for simplicity neglect the gluon spin term -- the last
term on the r.h.s. of (\ref{3}) (the  latter gives a correction to spatial
(magnetic) confinement), and then the gluon propagator can be written as
\be (-D^2)^{-1}_{xy}=<x|\int^{\infty}_0 dt e^{tD^2(B)}|y>=
\int^{\infty}_0dt(Dz)^w_{xy}e^{-K}\Phi(x,y) \label{5}\ee
 where
\be K=\frac{1}{4}\int^s_0d\tau \left(\frac{dz_\mu}{d\tau} \right)^2,
~~\Phi(x,y)=P \exp ig\int^x_yB_{\mu}dz_{\mu}\label{6}\ee
 and a winding path
measure is \be (Dz)^w_{xy}=\lim_{N\to
\infty}\prod^{N}_{m=1}\frac{d^4\zeta(m)}{(4\pi\varepsilon)^2}
\sum^{\infty}_{n=0,\pm1,..}
\int\frac{d^4p}{(2\pi)^4}e^{ip(\sum\zeta(m)-(x-y)-n\beta\delta_{\mu
4})}\label{7} \ee In the  free case, $B_\mu\equiv 0$, one obtains the gluon
propagator \be G(x,y)\to (
\partial^2)^{-1}_{xy}=\sum_{k=0,\pm1,...}\int \frac{T d^3p}{(2\pi)^3}
\frac{e^{-i\vec{p}(\vec{x}-\vec{y})-i2\pi kT(x_4-y_4)}}{(\vec{p}^2+ (2\pi
kT)^2)} .\label{8}\ee

At large distances the zero mode $(k=0)$ yields the behavior shown in
(\ref{3a}) and this   is the origin of the IR divergence of higher order $g^n$
contributions to the  free energy, as was shown in \cite{10}, while magnetic
confinement, contained in  $\Phi(x,y)$, cuts off all divergences, as  will be
demonstrated below.

One can easily find the lowest order (one loop) np contribution to the free
energy

\be F_0^{gl}(B)= T \left\{\frac{1}{2} \log \det G^{-1}-\log\det
(-D^2(B))\right\},\label{10} \ee which can be written as \be F_0^{gl} (B) =-T
\int^\infty_0 \frac{ds}{s} \xi (s) d^4 x (Dz)^w_{xx} e^{-K} \lan tr_a
\Phi(x,x)\ran_B\label{11}\ee and finally for the pressure $P_{gl}
=-\frac{1}{V_3} \lan F_0^{gl} (B)\ran$,

\be P_{gl}= (N^2_c-1) \int^\infty_0 \frac{ds}{s} \sum_{n\neq 0} G^{(n)} (s),
\label{12}\ee with \be G^{(n)} (s) =\int (Dz)^w_{on} e^{-K} \lan tr_a
\Phi(x,x)\ran_B\label{13}\ee  $\Phi(x,x)$ contains colorelectric fields
$B_4(x)$, which produce Polyakov lines $L_{adj} (T)$ \cite{3}, and in addition
also colormagnetic fields, which are  contained in the spatial Wilson loop,
$\lan tr_a\Phi_s ( C_n)\ran_B\equiv \lan W_s(C_n)\ran$, which can be written in
terms of field correlators \cite{1},  as an  integral over minimal surface
inside the loop $C$

\be \lan W_s(C_n)\ran = {tr_a } \lan \exp( ig \int_{C } A_\mu dz_\mu)\ran={tr_a
} \lan \exp( ig \int    ds_{\mu\nu} F_{\mu\nu})\ran\label{14}\ee and using the
cumulant expansion \cite{1,5,23} and dropping all cumulants except for
quadratic, one has

   \be \lan W_s(C_n)\ran =   \exp\left( -\frac{1}{2}
  \int_{S }\int_{S } d s_{\mu\nu} (u) d
  s_{\lambda\sigma} (v) \lan F_{\mu\nu} (u) F_{\lambda\sigma}
  (v)\ran \right).
  \label{15}
   \ee

   Considering only  spatial loops $C$ and surface areas $S$ for $k=0$, i.e. the term
   without higher Matsubara frequencies, one has to do with colormagnetic
   correlators only,
   \be \frac{g^2}{N_c}\lan H_i (u) H_j(v)\ran = \delta_{ij} D^H (u-v) + O(D^H_1).\label{16}\ee

To find $\sigma_s$ in (\ref{1a}) one can use the  connection of $D^H$ with  the
gluelump Green's function \cite{23}, which, as shown in the Appendix,
   can be written as \be D^H(z) = \frac{g^4
(N_c^2-1)}{2} T^2 G^{(2g)}_{3d} (z)   \label{21*}\ee where $G^{(2g)}_{3d}$ is
the two-gluon  np Green's function in 3d. As a result using (\ref{1a})  one can
write the $T$-dependent part of $\sigma_s$ as

\be \sigma_s (T)  = g^4 T^2 c^2_\sigma, ~~ c^2_\sigma =\frac{N^2_c-1}{4} \int
d^2 z G^{(2g)}_{3d} (z), \label{22*}\ee where $c_\sigma$ is dimensionless
number and a fully np quantity.

\begin{center}

\begin{figure}
\begin{center}
  \includegraphics[width=6cm, ]{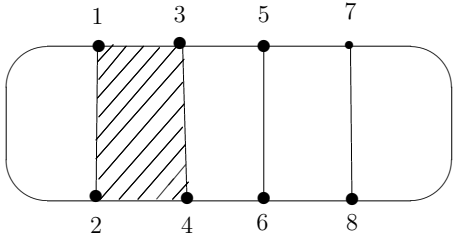}

\end{center}
  \caption{ The 8-th order graph with the crossed rectangular under study}

\end{figure}
 \end{center}

Insertion of  $\lan tr_a\Phi\ran$ in   (\ref{13})  as  an area law (\ref{1a})
yields a loop graph of a gluon, where the string tension $\sigma_s$ controls
the area inside the loop, so that the gluon cannot go far from the initial
point, the maximal distance being $R\la \frac{1}{\sqrt{\sigma_s}}$.
 One can now generalize this picture to the higher terms in the
perturbative series $O(g^n)$, where these terms are formed by applying the term
$L_3$ in the original QCD Lagrangian \be L_3 = g\partial_\mu a^a_\nu f^{abc}
a_\mu^b a_\nu^c\label{18}\ee on any gluon line. As a result one obtains e.g.
the diagram of  Fig.~1 of the order $g^8$. It is essential, that each gluon
propagator   $ G^a_{\mu\nu} (x^{(i)}, x^{(k)}) \equiv \lan a_\mu^a (x^{(i)})
a^b_\nu (x^{(k)})\ran$ is proportional to $\Phi(x^{(i)}, x^{(k)})$ and the
latter after averaging over background fields $B_\mu$, in the product together
with all other gluon propagators, forms the total Wilson loop with the same
outer contour $C_n$, but now with inner lines, dissecting it into a sum of
pieces of area   $\Delta A^{(i)}$, $\mathcal{A}\to \sum_i \Delta
\mathcal{A}^{(i)}$, each piece is subject to the area law with the same
$\sigma_s$, so that one obtains the  factor $\exp \left( - \sigma_s  \sum_i
\Delta \mathcal{A}^{(i)}\right)$, which prevents the escape of all gluons from
the  center of the area, and in this way ensures infrared stability.

One can say, that each gluon is  interacting with the closest neighbor via
linear confining interaction and therefore the distance between them is of the
order of $(\sqrt{\sigma_s})^{-1}$.

We now turn to the more formal procedure to define the properties of the
one-loop part of the complicated diagrams, shown in Fig.1, as a crossed
rectangular. At each vertex of this diagram  enters the operator (\ref{18}),
which generates 3g vertex $\Gamma_i$ with momentum operator $p_i$, so that the
quadratic  loop diagram in the 3d space can be written as

 \be G(p^{(1)}, p^{(2)}, p^{(3)}, p^{(4)}) =
\prod^4_{i=1} \Gamma_i \int^\infty_0  ds_i (Dz^{(i)})_{x^{(i)},x^{(i-1)}}
e^{-K_i} \Phi^{(i)} e^{ip^{(i)} x^{(i)}} dx^{(i)}.\label{31}\ee

Here we have introduced the phase factors \be \Phi^{(i)}(x^{(i)},x^{(i-1)})=
P_A \exp(ig \int^{x^{(i)}}_{x^{(i-1)}} A_\mu dz_\mu),\label{32} \ee omitting
for simplicity the gluon spin phase factor, originating from the last term in
(\ref{16}), since it is inessential in the asymptotic limit for large
$|x^{(i)}-x^{(i-1)}|.$ Here  $ (Dz^{(i)})$ is  \be (Dz^{(i)})_{xy}
=\lim_{N\to\infty}\prod^N_{k=1}
\frac{d^3\xi^{(i)}(k)}{(4\pi\varepsilon)^2}\frac{d^3q^{(i)}} {(2\pi)^4}
e^{iq^{(i)}(\sum_k\xi^{(i)}(k)-(x-y))},~~ N\varepsilon =s. \label{33} \ee
 It is  essential that the product of all phase factors $\Phi^{(i)}$ in the
 whole diagram of Fig.~1 should be averaged over vacuum configurations,
 yielding 3d confinement, and each gluon line is in  adjoint representation,
 and can be  represented as the double fundamental line in the simple case of
 the large $N_c$ limit, so that one finally has a product of independent closed
 fundamental lines, circumvented by  a common line in the outer contour.
  In  the  same large $N_c$ limit the average of this product
 can be   represented as the product of averages of individual loops time the
 average of the outer contour, which yields the overall confining factor. In
 what follows we shall be interested in the properties of one rectangular loop
 and demonstrate its spatial convergence, while the overall confining loop will
 exhibit additional convergence.

The rather complicated calculations, given in the Appendix B of the paper
 \cite{17} for the case $d=4$, can be done in an analogous manner for the case
$d=3$, and one obtains the following form  of the  rectangular   of Fig. 1 with
account of the spatial confinement

\be  G_4 (p_{ i})=(2\pi)^3 \delta^{(3)} \left(\sum p^{(i)}\right)
\prod^4_{i=1}\int \frac{d^3q_i \Gamma_i}{q^2_i}   I_4 (b),\label{37}\ee where
\be I_4 (b)  =  \int \frac{d^3\mathcal{P}}{(2\pi)^3}
\left(\frac{4\pi}{\sigma}\right)^6 \exp \left(-\frac{2}{\sigma}
\sqrt{b^2_1b_2^2- (\veb_1\veb_2)^2}\right)\exp \left(-\frac{2}{\sigma}
\sqrt{b^2_3b_4^2- (\veb_3\veb_4)^2}\right),\label{38}\ee and $b_i$ are

 $$ b_1=q_1-p_2-p_3+\mathcal{P},~~
b_2=q_2-p_3+\mathcal{P},$$ \be b_3=q_3+\mathcal{P},
~~b_4=q_4+p_4+\mathcal{P}.\label{39} \ee

 One can check, that at large momenta, (the hard regime)  when $b^2_i\gg
\sigma, ~i=1,2,3,4$. \be I_4 (b) \to \prod_{i=1,2,4} \delta^{(3)}
(b_i).\label{40}\ee and the product of four factors  $d^3q_i$ is reduced to a
single integration $d^3q_3$,
 as it   should
be in the free case without  confinement.

As a result one has in (\ref{37}) for  one loop in Fig.~1  the combination
$d^3q \prod^4_{i=1}\frac{\Gamma_i}{q^2_i}$, and for the whole chain of $n$
loops, as in Fig.~1, one  obtains an estimate (see \cite{10}) \be \mathcal{M}_n
(T) \sim g^{2(n-1)} \left( T \int^T_{a\sqrt{\sigma}} d^3 q\right)^n
\frac{q^{2(n-1)}}{(q^2)^{3(n-1)}}.\label{46a}\ee
 Here we have used the hard limit condition, $q\geq a \sqrt{\sigma}, a\gg 1$.

 Integration in (\ref{46a}) yields the result $(n>4)$
 \be \mathcal{M}_n^{\rm hard} (T) \sim \frac{g^{2(n-1)}  T^n
 \sigma_s^{\frac{4-n}{2}}}{a^{n-4}}\sim \frac{ g^6T^4}{(c_\sigma
 a)^{n-4}}\label{47a}\ee
where we have exploited (\ref{2a}). This, apart from the $c_\sigma a$ factor,
is the problem (2) of Linde \cite{10}: all terms with $n>4$ contribute to the
order $g^6T^4$, however with decreasing magnitude for $c_\sigma a \gg1$.

To complete our study we consider now the soft regime, all momenta $q_i, p_i$
in (\ref{39}) are small, $q_i, P \la \sqrt{\sigma}$. In this case every loop
integration $d^3q$ in (\ref{46a}) is replaced by

\be d^3q \to \prod^4_{i=1} d^3 q_i I_4 (b) = \sigma_s^{3/2} f\left(
\frac{q_i}{\sqrt{\sigma}}\right),\label{48a}\ee  where  $f$ in (\ref{38})
yields  a cut-off in the $d^3q$ integration in (\ref{46a}), and as a result one
obtains in the soft regime \be \mathcal{M}_n^{\rm soft} (T) \sim {g^{2(n-1)}
T^n
 \sigma_s^{\frac{4-n}{2}}}\varphi_n \sim   { g^6T^4}\varphi_n \label{49a}\ee
 where $\varphi_n$ is a converging integral of  dimensionless ratios  $q_i/
 \sqrt{\sigma}$. One can see, that (\ref{49a}) yields qualitatively the same
 result as in  (\ref{47a}),  for the order of magnitude
 estimates. However quantitatively one should calculate nonperturbatively the
 whole series $n\geq 4$ to recover the $O(g^6)$ contribution. This situation is
 similar to the solution of the relativistic  problem of two potentials: one confining and
 another gluon exchange but without small parameters, and  one should sum up the series, or rather solve the corresponding   relativistic
 Hamiltonian equation \cite{18}.

\section{Summary and prospectives}

We have considered above the gluon thermodynamics   with the nonperturbative
background fields, which ensure spatial confinement due to colormagnetic
correlators (\ref{16}). As a result one obtains the  area law of the spatial
Wilson loop with  the nonzero spatial string tension. Qualitatively it is
clear, that all multigluon diagrams in 3d would be convergent at large spatial
distances, and this property was used in \cite{12} to argue  that the Linde
problem is absent in the confining vacuum. In the present paper this
qualitative argument was given a more quantitative foundation.

Indeed, the explicit account of the spatial confinement  not only formally
solves the Linde problem, but it is  also vitally important in the
thermodynamics of the  quark gluon plasma (qgp), as was shown recently in the
case of  the SU(3) \cite{30,29}, as well as  in the case of $n_f =2+1$
thermodynamics in the deconfined phase \cite{30a}. It  was demonstrated there,
that  taking into account correlator $D^E_1$ (which  generates Polyakov lines)
and $D^H$ for spatial confinement one obtains a good agreement with most
accurate lattice data.

 As we argued above, qualitatively
  one can exploit the universal effective gluon mass $m^H_D(T) \cong
2 \sqrt{\sigma_s (T)}$  instead of  spatial confinement with $\sigma_s(T)$   as
a first approximation in the effective perturbation theory up to the $g^6$
order.

From this point of  view we have stressed the existence of the effective
screening mass parameter, which is of  np origin and occurs due to magnetic
confinement string tension $\sigma_s$ -- this is  the answer to  what we call
the problem  1 of Linde. The second problem of Linde, the infinite $g^n$ series
with $n\geq 6$,   contributing  to the order $g^6T^4$, is confirmed above in
the np approach.

As was mentioned above, the whole dynamics of  diagrams with $n\geq 6$ lies in
the soft np region, where the magnetic confinement and not gluon exchange
mechanism play  the most important role. It is an open question what is the sum
of $n>6$ np terms with magnetic confinement, which is equivalent to the $gg$
amplitude in the case of two interactions in 3d: confining $V_{\rm conf}$ and
gluon exchange $V_{OGE}$, but the answer is possible to obtain.

At this point it is worthwhile to compare our results with other approaches,
where the notion of confinement plays an important role. In the
Gribov-Zwanziger (GZ)    method (see e.g. \cite{13a,14a}) the local gluon mass
is introduced, which at large $T$ has the same form as the magnetic mass
$m_{\rm  mag} = c_m g^2 T$ in (\ref{a9}), but with  a different coefficient,
and  it it  is stated in \cite{14a}, that  the  resummation of the perturbation
series with this  mass yields results, consistent with lattice data.

It is not evident, that the  whole procedure of the gluon mass generation is
fully gauge invariant, however even assuming this, the simultaneous appearance
of confinement and the generated mass of gluons (and the use of only the second
factor) calls for additional inquiry.

Another  question of the  GZ approach is the possible  difficulty of the vector
confinement  with the so-called Klein paradox, leading to the  unboundedness of
the fermion spectrum in the confining vector potential,  see \cite{31} for
details.

Apart from this, the GZ approach suggests to solve the Linde problem 1 via
magnetic gluon mass generation.

Following this line of IR problem resolution via  confinement, it is worthwhile
to mention the active development of the methods of compactified QCD, and in
particular with respect to  weak-coupling  semiclassical realization of IR
renormalons as in \cite{32}. The problems of IR  renormalons have been also
studied in connection to the structure of singularities  in the Borel plane in
\cite{33}.

It is remarkable, how confinement   operates also in this respect. In \cite{34}
it was shown, that confinement resolves the IR renormalon problem, transforming
the series of renormalon loop graphs, covered with confining film, into the
Borel summable series. In this way the whole perturbative series in  the
confining background can be treated within the standard renormalization
technique without  IR renormalon singularities.

In a similar way the well-known Landau ghost problem of the IR divergence of
the renormalised $\alpha_s(Q^2)$ was solved  in \cite{35}. Summarizing, one can
say, that confinement actually solves  the known IR    problem of QCD.

 The
author is grateful for useful discussions to M.A.Andreichikov, B.O.Ker\-bi\-kov
and M.S.Lukashov. This work was done in the framework of the scientific
project, supported by the Russian Science Fund, grant 16-12-10414.

\setcounter{equation}{0} \def\theequation{A1.\arabic{equation}}

\vspace{2cm}
 \setcounter{equation}{0}
\renewcommand{\theequation}{A.\arabic{equation}}

\hfill {\it  Appendix  1}

\centerline{\it\large Calculation of the spatial string  tension via  two-gluon
 Green's function  }

 \vspace{1cm}

\setcounter{equation}{0} \def\theequation{A1.\arabic{equation}}

To calculate $D^H(z)$ one can use the technic, developed in \cite{23} for
$D^E(z)$, which allows to express it via two-gluon  Green's function
$G^{(2g)}_{4d} (z) =  G_{4d}^{(g)}\otimes G_{4d}^{(g)}$, where two gluons
interact nonperturbatively.

The starting point  for the gluon propagator $G^{(g)}_{4d}$ is the integration
in the 4 th direction in (\ref{5}) with the exponent $K_4 = \frac14 \int^s_0 d
\tau \left( \frac{d z_4}{d\tau}\right)^2,$ which gives for the spatial loop
with $x_4=y_4$, \be J_4 \equiv \int (Dz_4)_{x_4x_4} e^{-K_4} = \sum_{n=0, \pm
1,...} \frac{1}{2 \sqrt{\pi s}} e^{-\frac{(n\beta)^2}{4s}} =
\frac{1}{2\sqrt{\pi s}} \left( 1+ \sum_{n=\pm 1,\pm 2}
e^{-\frac{(n\beta)^2}{4s}}\right).\label{*}\ee

The  second term in (\ref{*}) at large $T\gg \frac{1}{2\sqrt{s}}$ yields
$2\sqrt{\pi s} T$,    which gives $J_4 =\frac{1}{2\sqrt{\pi s}}+T$.

The same linear in $T$ term is obtained using the Poisson relation \cite{2,3}.
As  a result the 4d gluon propagator is reduced to the  3d one, \be
G_{4d}^{(g)}(z) = T G_{3d}^{(g)}(z) + K_{3d} (z)\label{**}\ee where $K_{3d}
(z)$ does not depend on $T$. In what follows we consider only the first term in
(\ref{**}),  keeping in mind, that  $G_{4d}^{(g)}(z)$ at small $T$ has  a
nonzero limit.  Substituting this term in  the  general expression for $D^E(z)$
$(D^H(z))$ obtained in \cite{23}, one has

 \be D^H(z) =
\frac{g^4(N^2_c-1) }{2} \lan G^{(2g)}_{4d} (z)\ran  \to
\frac{g^4(N^2_c-1)T^2}{2} \lan G^{(2g)}_{3d} (z)\ran, \label{A.1}\ee where $
G^{(2g)}_{3d}$ is the two-gluon Green's function in 3d with all interaction
between gluons taken into account \be\lan  G^{(2g)}_{3d}\ran = \lan
G^{(g)}_{3d}(x,y) G^{(g)}_{3d}(x,y)\ran_B.\label{A.2}\ee In terms of the
gluelump phenomenology, studied in \cite{17*,24} (\ref{A.2}) is called the
two-gluon gluelump, computed on the lattice in \cite{24} and analytically in
\cite{17*}. In our case we are interested in the 3d version of the
corresponding Green's function. Choosing in 3d the $x_3\equiv t$ axis as the
Euclidean time, we proceed as in \cite{7}, exploiting the path integral technic
\cite{1,17}, which yields \be G^{(2g)}_{3d} (x-y) = \frac{t}{8\pi}
\int^\infty_0 \frac{d\omega_1}{\omega_1^{3/2}} \int^\infty_0
\frac{d\omega_2}{\omega_2^{3/2}} (D^2z_1)_{xy}(D^2z_2)_{xy}
e^{-K_1(\omega_1)-K_2(\omega_2)-Vt},\label{A.3}\ee where $V$ includes spatial
confining interaction between the three objects: gluon 1, gluon 2 and  the
fixed straight line of the parallel transporter, which makes all construction
gauge invariant (see \cite{17*,23} for details). In (\ref{A.3}) $t=|x-y|\equiv
|w|;$ and finally \be \sigma_s (T) =\frac{g^4(N^2_c-1)T^2}{4} \int \lan
G^{(2g)}_{3d} (w)\ran d^2w.\label{A.4}\ee Constructing in the exponent of
(\ref{A.3}) the 3 body Hamiltonian in the 2d spatial coordinates \be
H(\omega_1, \omega_2) = \frac{ \omega_1^2+ \vep^2_1}{2\omega_1}+  \frac{
\omega_2^2+ \vep^2_2}{2\omega_2}+ V(\vez_1, \vez_2), \label{A.5}\ee one can
rewrite (\ref{A.3}) as follows, see \cite{17}  \be G^{(2g)}_{3d} (t) =
\frac{t}{8\pi} \int^\infty_0 \frac{d\omega_1}{\omega_1^{3/2}} \int^\infty_0
\frac{d\omega_2}{\omega_2^{3/2}} \sum^\infty_{n=0} |\psi_n (0,0)|^2 e^{-M_n
(\omega_1,\omega_2) t}.\label{A.6}\ee Here $\Psi_n(0,0)\equiv \Psi_n (\vez_1,
\vez_2)|_{\vez_1=\vez_2=0}$, and $M_n$  is the eigenvalue of $H(\omega_1,
\omega_2)$. The latter was studied in \cite{17*} in 3 spatial coordinates. For
our purpose here we only mention, that $ G^{(2g)}_{3d}  (z)$ has the dimension
of the mass squared and the integral in (\ref{A.4})  is therefore
dimensionless. Hence one obtains $\sqrt{\sigma_s (T)} = g^2 T  c_\sigma
+$~const, as was stated above in (\ref{3a}), where \be  c^2_\sigma =
\frac{(N_c^2-1)}{4} \int d^2 w \lan G_{3d}^{(2g)} (w)\ran, \label{A.17}\ee and
 const is obtained from the second term in (\ref{**}).

\end{document}